\documentclass[aps,prb,twocolumn,floatfix,superscriptaddress]{revtex4}
\usepackage[dvips]{graphicx}

\begin{document}

\title{Effect of quantization of vibrations on the structural properties of crystals}


\author{Iv\'an Scivetti}
\affiliation{Atomistic Simulation Centre, Queen's University Belfast, Northern Ireland, UK}
\author{Nikitas Gidopoulos}
\affiliation{ISIS Facility, Rutherford Appleton Laboratory, Chilton, Didcot, England, UK}
\author{Jorge Kohanoff}
\email[]{j.kohanoff@qub.ac.uk}
\affiliation{Atomistic Simulation Centre, Queen's University Belfast, Northern Ireland, UK}

\begin{abstract}
We study the structural effects produced by the quantization of vibrational degrees
of freedom in periodic crystals at zero temperature. To this end we introduce a 
methodology based on mapping a suitable subspace of the vibrational manifold and 
solving the Schr\"odinger equation in it. A number of increasingly accurate approximations 
ranging from the quasi-harmonic approximation (QHA) to the vibrational self-consistent 
field (VSCF) method and the exact solution are described. A thorough
analysis of the approximations is presented for model monoatomic and hydrogen-bonded chains,
and results are presented for a linear HF chain where the potential energy surface is
obtained via first-principles electronic structure calculations. We focus on quantum nuclear 
effects on the lattice constant, and show that the VSCF is an
excellent approximation, meaning that correlation between modes is not extremely important.
The QHA is excellent for covalently-bonded, mildly anharmonic systems, but it fails
for hydrogen-bonded ones. In the latter, the zero-point energy exhibits a non-analytic
behavior at the lattice constant where the H-atoms center, which leads to a spurious 
secondary minimum in the quantum-corrected energy curve. An inexpensive anharmonic 
appoximation of non-interacting modes appears to produce rather good results for
hydrogen-bonded chains, for small system sizes. However, it converges to the incorrect QHA 
results for increasing size. Isotope effects are studied for the first-principles HF
chain. We show how the lattice constant and the HF distance increase with decreasing 
mass, and how the QHA proves to be insufficient to reproduce this behavior.
\end{abstract}

\date{\today}
\maketitle


\section{Introduction}

Structural properties of solids like interatomic distances, bond angles and equilibrium 
lattice parameters are customarily calculated by assuming that atomic nuclei (or ionic 
cores) behave as classical particles. Within this framework, the electronic problem is 
solved for a fixed configuration of clamped nuclei. The resulting ground state electronic 
density allows for the computation of the forces acting on the nuclei through 
Hellman-Feynman's theorem. These forces are then used to determine 
the equilibrium configuration. While this is justified for a large
variety of systems of interest, the assumption of classical nuclei becomes questionable 
whenever light atoms such as hydrogen are involved. Moreover, this approximation can not 
address isotope effects; the ground state electronic energy does not depend on the 
nuclear masses but only on the atomic numbers. A widely-used approach to introduce the 
quantum nature of the nuclei is the quasi-harmonic approximation (QHA).\cite{qha1,qha2,qha3} In this 
approximation, the nuclear ground state energy is supplemented with the zero-point-energy 
(ZPE) corresponding to {\it harmonic} nuclear vibrations. This is based on a second-order 
Taylor expansion of the potential energy surface (PES) in terms of the atomic coordinates 
around the equilibrium configuration of the nuclei. The resulting dynamical matrix, given 
by the second derivatives of the potential with respect to the nuclear coordinates, can 
be diagonalized thus leading to a set of orthogonal eigenvectors or normal modes 
$\zeta_l$, together with the corresponding set of frequencies $\omega_l$.\cite{kittel}

To simulate an infinite crystal, it is customary to apply periodic boundary conditions 
on the unit cell.\cite{kittel} Therefore, a crystal described by a basis of $N$ atoms 
in the unit cell is characterized by $3N$ vibrational bands $\omega_l({\bf k})$, where 
${\bf k}$ is a wave vector in the phonon Brillouin zone (BZ). Within this framework, the QHA 
energy at zero temperature, $E^{QHA}$, is defined as
\begin{equation}
E^{QHA}(V)=E^{cl}(V)+\frac{\hbar}{2}~\sum_{l=1}^{3N} 
\int_{BZ} g({\bf k})~\omega_l({\bf k},V)\,d{\bf k}
\label{e_qha}
\end{equation}
where $V$ is the volume of the system, $g({\bf k})$ is the density of states and the 
integral extends to the Brillouin zone. The first term in the RHS is the ground 
state energy for the equilibrium configuration at volume $V$. The second 
term includes the quantum nature of the nuclei through the harmonic ZPE, which also 
depends on $V$ in a way that is characteristic of the type of bonding. For covalent 
bonding it increases upon compression, while for hydrogen-bonding there is a competition 
between increasing and decreasing frequencies. The combination of ground-state and 
zero-point energies leads to a modified $E^{QHA}$ vs. $V$ curve -- in contrast 
to $E^{cl}$ vs. $V$, with a minimum located at a volume corrected by quantum nuclear 
effects. The QHA has the additional advantage of lending itself naturally to the 
simultaneous incorporation of quantum and thermal effects. \cite{scivetti_aip}

In the harmonic approximation the ground state nuclear wave function is a product of 
single-mode Gaussians on the vibrational normal mode variables $\zeta_l$, and the 
expectation value of the nuclear coordinates coincides exactly with the classical nuclear 
configuration. Apart from variations mediated by volume changes, internal structural 
parameters (distances and angles) are unaffected in the QHA. It is then expected that 
the QHA breaks down when treating highly anharmonic systems.  

Additional terms in the Taylor expansion lead to anharmonicity, which can come 
essentially in two forms. The simplest one is {\it intra-mode} anharmonicity, 
when the vibrational modes remain non-interacting but the potential felt by one 
or more modes cannot be approximated by a quadratic expression. This type can be 
included in a rather simple way. Normal modes can be used as coordinates to map
the PES beyond the harmonic level, and the energy of each mode obtained by solving 
a set of independent one-dimensional Schr\"odinger equations.
The second type of anharmonicity is due to mode-coupling. When the excursions along 
a vibrational normal coordinate are large, the approximation of small oscillations 
implicit in the second-order expansion breaks down. Successive terms in the expansion 
involve products of modes of the form $\zeta_l^m \zeta_{l^{\prime}}^n$ as well as
higher-order terms involving more than two modes. Which groups of modes are more
strongly coupled depends very much on the specific system. 
In general there are no rules to find out a priori which couplings need to 
be considered. A possible strategy is to displace the system along one mode, say 
$\zeta_1$, by an amount $\epsilon$, and then optimize the atomic coordinates under 
the constraint $\zeta_1=\epsilon$. By projecting the displacements obtained for the 
optimized configuration onto the original normal modes, one can identify and select 
the modes with the largest projection.\cite{giuseppe_thesis} The main goal of this 
paper is to evaluate the quality of the QHA and, where required, to provide improved 
schemes at an affordable computational cost. To this end, we will include anharmonicity 
by solving the vibrational Schr\"odinger equation in a sequence of increasingly 
accurate approximations, and compare to the QHA results.

An additional issue that arises when computing the ZPE is that the BZ integrals in 
(\ref{e_qha}) must be replaced with appropriate averages over a finite set of
representative points in the BZ. This can be done in several 
ways.  The most accurate one is to compute the force constants in real space and 
then use them to obtain the dynamical matrix at an arbitrarily dense mesh of 
${\bf k}$-points in the BZ. This requires calculations in large supercells or,
alternatively, linear response calculations. \cite{baroni_01} On the other extreme, 
the crudest approximation is to include only the BZ-center modes of the unit cell,
while improvements can be achieved by considering BZ-center modes of larger supercells. 
The advantage of this latter is that it is easier to extend to anharmonic 
situations. A second goal of this paper is to analyze how large a supercell 
should be in order to reproduce the equilibrium structure obtained with converged 
BZ averaging. This issue has rarely been discussed in the literature. 

In order to answer these questions we have studied two qualitatively different models: 
a one-dimensional monoatomic chain and a diatomic linear chain that models a 
double-well hydrogen-bonded system. In Section \ref{monoatomic} we show that, as
usually done in the literature, systems characterized by covalent bonding can be
safely described within the quasi-harmonic approximation using a rather coarse
BZ sampling. In Section \ref{diatomic}, in contrast, we show that hydrogen-bonded 
systems require some level of anharmonicity in their treatment. An anomaly in the 
ZPE appears when the protons center in the H-bonds, upon compression. This produces
an unusual behavior of the QHA, which can lead to a secondary spurious minimum in
the energy-volume curve. We then study a realistic linear F-H$\cdots$F chain where the 
PES has been obtained from first-principles calculations (Section \ref{first_principles}). 
We study quantum effects, in particular the isotope effect on equilibrium lattice constant 
and internal geometry (H-F distance) occurring when the nuclear masses are modified
(Section \ref{isotope}). In this case, an inexpensive anharmonic approximation that 
neglects mode-coupling appears to produce an accurate lattice constant. In Section 
\ref{conclusions} we present our conclusions and elaborate on possible extensions. 
In the following, we introduce the theoretical and computational approaches used 
in the present work.

\section{Methods and approximations}
In this paper we will only consider one-dimensional systems, for the sake of
simplicity. Although computationally more demanding, extensions to higher dimensions 
are straightforward. In the QHA, the energy for a one-dimensional system containing
$N$ particles is
\begin{equation}
E^{QHA}(a)=E^{cl}(a)+\frac{\hbar}{2}~\sum_{i=1}^{N-1}~\omega_i(a)~~,
\end{equation}
where $a$ is the lattice constant -- which plays the role of the volume, and $E^{cl}(a)$ 
is the ground state eletronic energy in the approximation of classical nuclei (classical energy). 
The vibrational frequencies $\omega_i(a)$ are 
obtained by assuming periodic boundary conditions (PBC), i.e. we consider a cyclic chain. 
The sum runs up to $N-1$ because the vibrational mode corresponding to the rigid translation 
of the whole chain has zero frequency. 

In the harmonic approximation there is no need to consider explicitly the $N$-atom chain. The same 
result can be obtained by studying monoatomic or diatomic cells subject to PBC. This 
equivalence is explained thoroughly in most solid state books. \cite{kittel} 
Here we just quote the main result, i.e. the dispersion relations that express the 
frequency as a function of wavenumber $k$ in the one-dimensional phonon Brillouin zone. 
For a monoatomic chain the dispersion relation is 
\begin{equation}
\omega(k,a)=2\sqrt{\frac{V^{\prime\prime}(a)}{m}}~
\left|\sin\left(\frac{ka}{2}\right)\right|~~,
\label{omega}
\end{equation}
with $m$ the mass of the particles, $k=\pi n/(N-1)a$ and $n=0,1,\cdots,N-1$. 
The quantity $V^{\prime\prime}(a)$ is the second derivative (curvature) of the 
interaction potential $V$ evaluated at the equilibrium configuration, and depends 
on the lattice constant $a$.
The frequencies obtained in this way correspond exactly with those of an $N$-atom 
cyclic chain. In the limit of $N\to \infty$, the QHA energy for a monoatomic chain is
\begin{equation}
E^{QHA}(a)=E^{cl}(a)+\frac{\hbar}{2}\int_{BZ} g(k)\,\omega(k,a)\,dk~~.\label{e_qha_1d}
\end{equation}
The case of a diatomic linear chain is analogous, with the difference that there are
two dispersion relations, representing the acoustic and optical branches. The expressions
are still analytic, but slightly more complicated. \cite{kittel}

For a hydrogen-bonded diatomic chain like F-H$\cdots$F the harmonic approximation can only 
be carried out at one of the two equivalent global minima. If the barrier is so high that 
tunnelling is unlikely, this is a reasonable approximation. However, if tunnelling
is important, anharmonic effects start to play a significant role and the QHA breaks down. 
By symmetry, one could also choose a reference configuration at the top of the inversion 
barrier, i.e. with the H-atoms centered between F-atoms. This configuration is 
a stationary point of the PES, but the QHA breaks down because a whole portion of the 
optical phonon branch is unstable around zone-center. 

The calculated normal modes can now be used to map the true PES, beyond the harmonic approximation. 
In general, this is a complicated mutidimensional fitting problem that requires specific 
techniques such as the product representation \cite{product}. A much simpler alternative is 
to keep considering the normal coordinates as non-interacting, but anharmonic. This
corresponds to pure intra-mode anharmonicity. The full-dimensional vibrational problem
is then reduced to mapping the one-dimensional potential for every normal coordinate
keeping the other normal coordinates at their classical value, i.e. zero. Then, the 1-D 
Schr\"odinger equation is solved for each one of these potentials and their ground state 
energies are added up. We call this the {\it anharmonic} approximation (ANHA).


At the other end of the theoretical spectrum, the problem of interacting phonons 
can be tackled exactly by solving the $(N-1)$-dimensional Schr\"odinger equation. 
This task can be accomplished with little effort for three degrees of freedom, but 
finds a hard wall at six. This means that, in the one-dimensional case, one can 
routinely compute the exact solution for a 4-atom supercell, but cannot go 
beyond 7-atom chains if a significant fraction of the eigenvalue spectrum is required. 
Stochastic methods such as diffusion Monte Carlo can, however, be used to compute 
the low-energy region of the spectrum, as it has been done for vibrations
in molecular systems. \cite{watts}

Here we have chosen the strategy of solving the vibrational Schr\"odinger equation 
using a discrete variable representation. We used Lagrange grids based on a 
combination of Cartesian and Hermite orthogonal polynomials \cite{baye,varga} for 
double-well and single-well potentials, respectively. When solving the Schr\"odinger
equation exactly, a maximum of three vibrational degrees of freedom was considered. 
This implies that the largest monoatomic linear chain we studied was made of four 
unit cells, while the largest diatomic chain contained two unit cells.

We have also studied a class of approximations to the full-dimensional solution
inspired by the vibrational self-consistent field (VSCF) method, \cite{gerber} which
is frequently used to compute vibrational excitation spectra of large molecules.
\cite{adrian} In the VSCF method the normal coordinates are taken as
completely uncorrelated and the total wave function is written 
as a product of single-mode wave functions:
\begin{equation}\label{hartree_prod}
\Psi(\zeta_1,\cdots,\zeta_{N-1})=\prod_{i=1}^{N-1}\phi_i(\zeta_i)\,\,,
\end{equation}
in the same spirit of the Hartree approximation to the problem of interacting
electrons. Contrary to the harmonic and anharmonic approximations, each mode now
feels the presence of the other modes, but in a mean-field way. The
VSCF equations are
\begin{equation}
\left[-\frac{\hbar^2}{2M}\frac{d^2}{d\zeta_i^2}+V_i(\zeta_i)\right]\phi_i(\zeta_i)=
\epsilon_i\,\phi_i(\zeta_i)~~,
\end{equation}
where
\begin{equation}
V_i(\zeta_i)=\int\cdots\int\, V(\zeta_1,\cdots,\zeta_{N-1})
\prod_{j=1,j\ne i}^{N-1} \mid\phi_j(\zeta_j)\mid^2d\zeta_j
\label{pot_vscf}
\end{equation}
are the mean-field potentials for each mode. For a 1-D system the energy in the VSCF 
method is given by
\begin{equation}
E_{VSCF}=\sum_{i=1}^{N-1}\,\epsilon_i - 
(N-2)\int V_i(\zeta)\,\mid\phi_i(\zeta)\mid^2d\zeta
\end{equation}
where the second term corrects for double-counting in the sum of eigenvalues. Notice
that the integral in this term should be independent of $i$, within numerical accuracy.

The determination of the mean-field potential requires the computation of 
$(N-2)$-dimensional integrals, which rapidly becomes an expensive operation. 
Therefore, it is desirable to find either simpler approximations or some
way of evaluating these integrals at a reduced computational cost.
The {\it anharmonic} approximation (ANHA) mentioned above removes this limitation 
by replacing the single-mode densities $|\phi_j(\zeta_j)|^2$ in (\ref{pot_vscf}) 
with delta-functions centered at $\zeta_j=0$. With this, the ANHA can be carried 
onto much larger systems.

The quality of the VSCF results depends on the choice of coordinates. If the system is 
only weakly anharmonic, then the classical normal modes are a very reasonable choice. 
This is not necessarily the case when anharmonicity becomes important. The question 
arises of whether it is still possible to find a set of orthogonal coordinates that 
are only weakly correlated, or whether correlation is intrinsic to the problem and
cannot be significantly reduced by a clever choice of coordinates.
Within the VSCF context, finding the optimally de-correlated coordinates is equivalent 
to requiring that these minimize $E_{VSCF}$. The optimal coordinates $\{\zeta_i\}$ can 
be obtained by solving the VSCF problem for a linear combination of normal modes 
$\zeta_i^{\prime}=\sum_j\,Z_{ij}\,\zeta_j$, and minimising $E_{VSCF}$ with respect to the 
coefficients $Z_{ij}$ of the rotation matrix, subject to orthonormality constraints. 
\cite{truhlar} This task can be accomplished by constrained multidimensional 
optimisation algorithms. \cite{gallegos} Nevertheless, there is no clear evidence 
of a general method for obtaining the optimal set of coordinates. In this work we 
analyze the quality of the VSCF approximation with respect to the choice of vibrational 
coordinates for the ab initio hydrogen-bonded linear chain. 

\section{Results}

\subsection{Model monoatomic chain\label{monoatomic}}

We first considered a periodic monoatomic chain where the atoms interact via a Morse potential
of the form
\begin{equation}
V(x)=D\left(1-e^{-b\,(x-a_0)}\right)^2\,\,,
\label{morse}
\end{equation}
where parameter $D$ represents the binding energy of a dimer, $a_0$ is the location 
of the minimum of the potential, which corresponds to the classical lattice constant,
and $b$ is a parameter that determines the curvature of the potential through the relation
$V''(a_0)=2Db^2$. The Morse potential is intrinsically anharmonic, and allows us to study 
the effect of quantization of the vibrations on the lattice constant. For the various 
approximations described in the previous Section we calculated the quantum-corrected 
energy as a function of the lattice parameter $a$, and then determined the minimum of the 
$E$ vs. $a$ curve.

In the solution of Schr\"odinger's equation, the extent of quantum nuclear effects can be 
measured by the product $\gamma=mD$, with $m$ the mass of the particles and 
$D$ the energy scale of the potential. Small values of $\gamma$ lead to important quantum 
effects, becoming less relevant as $\gamma$ increases, and eventually converging towards 
the classical results for $\gamma\to\infty$. Therefore, we report our results in terms of 
this quantity. In this work we have used $a_0=1.3983$ Bohr and $b=2.9864$ Bohr$^{-1}$.

\begin{figure}[ht]
\centering
  \includegraphics[scale=0.3,angle=270]{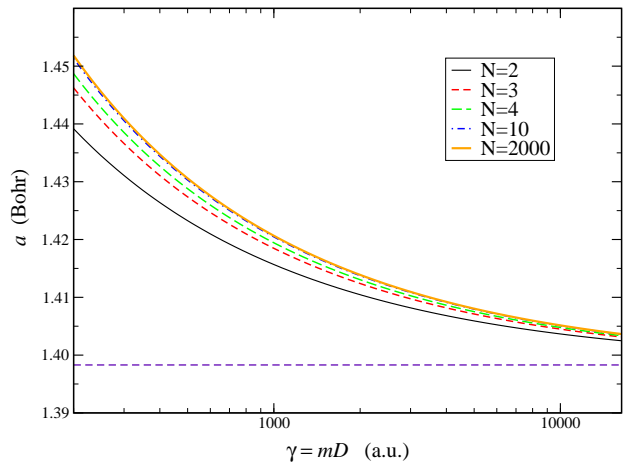}
  \caption{(Color online) QHA lattice constant as a function of $\gamma$, for 
           increasing number of atoms in the supercell: $N=2$ (black, thin solid), 
           $N=3$ (red, short-dashed), $N=4$ (green, long-dashed), $N=10$ (blue,
           dot-dashed), and $N=2000$ (orange, thick solid-light grey). The dashed 
           horizontal line represents the classical value.}
  \label{a_vs_m_mon}
\end{figure}

We first analyzed the effect of Brillouin zone sampling on the equilibrium lattice 
constant. For the QHA this can be done for an arbitrarily large number $N$ of k-points 
in the BZ, which is equivalent to considering a supercell containing an equal number 
$N$ of unit cells. In Fig. \ref{a_vs_m_mon} we show the convergence pattern of the 
QHA lattice constant as a function of $\gamma$, for increasing values of $N$. 

It is obvious that quantum effects are more important for smaller masses and for 
weaker interaction potentials. 
For H-atoms and $D=2$ eV ($\gamma=135$ a.u.) we are in the region to the left of 
Fig.\ref{a_vs_m_mon}, where the 
quantum-corrected lattice constant is about 4\% larger than the classical value 
(horizontal line). It can be seen that, for this type of potential, 10 cells
are already sufficient to reproduce the infinite
crystal lattice constant to high accuracy. However, 4 cells already 
produce excellent results and 2 cells underestimate $a$ in less that 1\%.

\begin{figure}[ht]
  \centering
  \includegraphics[scale=0.3,angle=270]{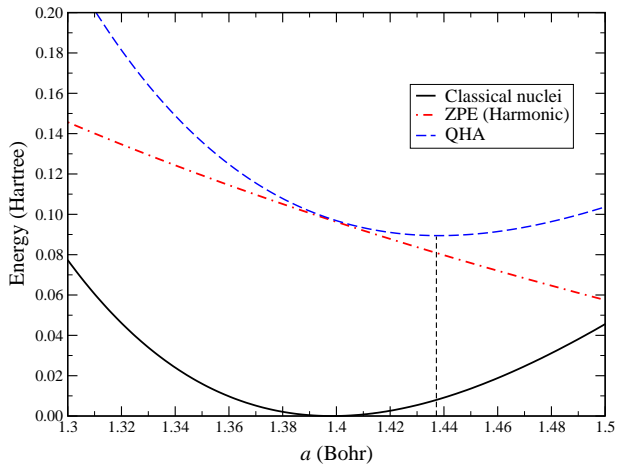}
  \caption{(Color online) Contribution of the harmonic ZPE as a function the lattice 
           parameter for a 4-atom supercell and $\gamma$=304.97 a.u. Since the ZPE 
           increases as $a$ decreases (red, dot-dashed line), the QHA lattice 
           parameter (vertical dashed line) corresponding to the minimum of the QHA 
           energy curve (blue, dashed), is expanded.}
\label{zpe_harmonic_vs_a_monoatomic}
\end{figure}

In order to illustrate the origin of the lattice expansion, we show in Fig.  
\ref{zpe_harmonic_vs_a_monoatomic} the various energy curves for a 4-atom supercell, 
as a function of the lattice parameter for $\gamma$=304.97 a.u. The classical energy 
curve $E^{cl}(a)$ (black, solid line) exhibits a minimum at the equilibrium lattice 
constant, where the one-dimensional pressure, $P=-dE^{cl}(a)/da$, vanishes. 
The ZPE (red, dot-dashed line) depends on the lattice constant through the curvature 
of the potential. Upon compression (reducing $a$), the chain becomes stiffer, all 
frequencies increase and so does the ZPE. Therefore, the QHA energy given by the 
sum of the two terms in (\ref{e_qha_1d}), and represented by the blue, dashed curve, 
exhibits a minimum at an expanded lattice constant. This effect is 
similar to thermal expansion, but the origin is purely quantum-mechanical. It can also
be ascribed to a zero-point pressure originated in the dependence of the ZPE on the
lattice constant.

\begin{figure}[ht]
  \centering
  \includegraphics[scale=0.3,angle=270]{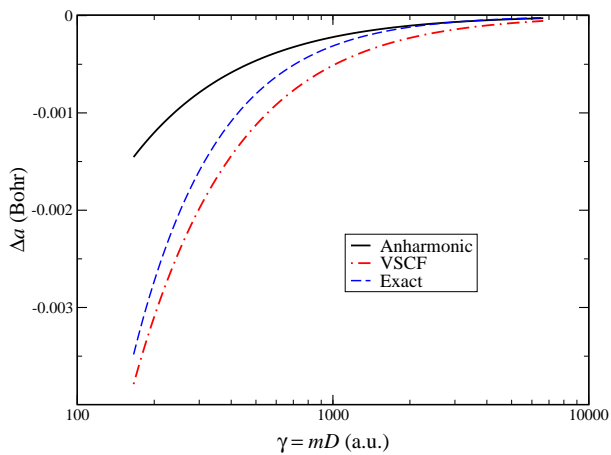}
  \caption{(Color online) Difference in equilibrium lattice constant with respect to 
           the QHA value, as a function $\gamma$ in a 4-atom ceck ll. The black
           solid curve is for the ANHA, the red (dot-dashed) for the VSCF, and 
           the blue (long-dashed) for the exact calculation.}
  \label{comp_meth_mon}
\end{figure}
To evaluate the quality of the QHA we calculated the equilibrium lattice constant in 
the ANHA and VSCF approximations, and exactly for a 4-atom supercell. The differences 
between these results and the QHA lattice parameter, $\Delta a$, are reported in 
Fig. \ref{comp_meth_mon}.
It can be seen that these differences are one order of magnitude smaller
than those arising from a poor sampling of the BZ. 
Thes observations justify the approach commonly used to determine the quantum-corrected
lattice constant of solids, i.e. that of computing the harmonic ZPE for a coarse 
BZ sampling -- preferably only the $\Gamma$-point --, adding it to the electronic ground state 
energy corresponding to the equilibrium configuration, and obtaining the lattice constant as the minimum of the sum. As we shall
see now, however, this procedure is not necessarily useful for very anharmonic 
systems such as hydrogen-bonded crystals.

\subsection{Model hydrogen-bonded chain\label{diatomic}}

As an extension to highly anharmonic systems, we considered a model for a 
hydrogen-bonded diatomic chain where the light atoms are generally hydrogens, and 
the heavy atoms can be oxygen or fluorine. For the sake of simplicity, and to be 
consistent with the following Section, we call them H and F, respectively. For the 
F-H potential we retain a Morse-type of interaction (Eq. \ref{morse}). For sufficiently 
large F-F distances the H atoms feel attracted to one or the other F-atom, rather than 
being shared between them. Therefore, the H-atoms in the chain are subject to a 
double-well potential. This potential has an important property for what concerns 
isotope effects: the barrier between the two wells becomes lower when the two 
neighbouring F-atoms approach each other. This is an essential ingredient, which 
allows for an easier migration of the protons upon compression. The F displacements 
($\rho_{i}$) are measured from the given lattice parameter. On the other hand, 
the H displacements ($u_{i}$) are measured from the midpoint between the 
equilibrium positions of the F-atoms. Displacements $\rho_{i}$ and $u_{i}$ are continuous 
variables associated to discrete lattice sites $i$. With this, the double Morse 
contribution to the PES is
\begin{eqnarray}\label{pes_fh}
{\cal E}_{FH}=
\sum_{i=1}^{N}&D&\left(1-e^{-b\left(\frac{a}{2}+u_i-\rho_i-r_e\right)}\right)^{2}\nonumber\\
+\sum_{i=1}^{N-1}&D&\left(1-e^{-b\left(\frac{a}{2}-u_i+\rho_{i+1}-r_e\right)}\right)^{2}\nonumber\\
+&D&\left(1-e^{-b\left(\frac{a}{2}+u_N-\rho_1-r_e\right)}\right)^{2}~~,
\end{eqnarray}
where $N$ is the number of sites and $a$ is the lattice parameter. The potential parameters 
are the well depth $D$ and the position of the minima $r_{e}$, measured from the F-atoms. 
The effects of periodic boundary conditions (PBC) are included in the last term on the RHS.

When the H-atom sits off-center it forms a neutral unit with the F-atom, and another H-atom 
on the opposite side of the F is not welcome due to electronic closed-shell repulsion effects. 
A similar effect occurs in ice, where oxygens make two strong covalent bonds with H-atoms to 
form the water molecules, and two weaker hydrogen bonds with neighboring molecules. This, 
however, is not taken into account by the Morse potential. To introduce this feature, usually 
known as Pauling's ``ice-rules", it is necessary to include an interaction between the 
H-atoms that discourages them from approaching the same F-atom. There are several possible 
ways of doing this. A popular choice is to include a harmonic interaction between H-atoms. 
The H-H contribution to the PES is
\begin{equation}\label{pes_hh}
{\cal E}_{HH}=\sum_{i=1}^{N-1}k_{HH}\left(u_{i+1}-u_{i}\right)^{2}+
k_{HH}\left(u_{1}-u_{N}\right)^{2}~~,
\end{equation}
where the last term takes care of PBC. This potential encourages adjacent H-atoms to move 
the same amount in the same direction, thus promoting the ice rules. The larger the 
constant $k_{HH}$, the more effectively these are respected. A similar term is included 
for the F-atoms
\begin{equation}\label{pes_ff}
{\cal E}_{FF}=\sum_{i=1}^{N-1}k_{FF}\left(\rho_{i+1}-\rho_{i}\right)^{2}+
k_{FF}\left(\rho_{1}-\rho_{N}\right)^{2}~~,
\end{equation}
with the corresponding PBC. This potential favors the F-atoms to move in the same direction so 
that two F-atoms cannot approach the same H at once. This model has been taken from a work 
by Yanovitskii et al, \cite{yanovitskii} who studied the phase diagram in the self-consistent 
harmonic approximation for a O-H$\cdots$O linear chain. The only difference with respect to 
Yanovitskii's model is the value of the harmonic constant $k_{HH}$ (0.06413 instead of 0.01413
a.u.). This modification was necessary because the original value did not enforce the ice 
rules strongly enough, thus leading to a potential with quite unrealistic features such as 
unstable acoustic phonon branches, as shown in a preliminary version of this work.
\cite{scivetti_aip}

\begin{figure}[ht]
  \includegraphics[height=.15\textheight]{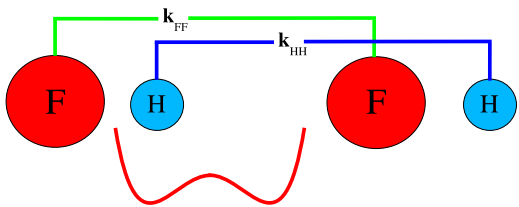}
  \caption{(Color online) Schematic view of the interactions in a model for a hydrogen-bonded chain.}\label{model_hf}
\end{figure}

This model assumes that the H-H and the F-F interactions are independent of the 
lattice parameter $a$. 
A mild dependence on the lattice constant enters through the H-F interaction, but it is quite
unrealistic. In order to render quantum effects more realistic we have introduced an additional 
Morse-type potential
\begin{eqnarray}\label{pes_a}
{\cal E}_{a}=\sum_{i=1}^{N}D_l\left(1-e^{-b_l\left(a-a_l\right)}\right)^{2}~~.
\end{eqnarray}
The parameters used in the present work were inspired in the calculations presented in the
following Section, and are summarized in Table \ref{param}. With this choice of parameters,
the classical lattice constant is $a_{cl}=4.368$ Bohr. In Fig. \ref{model_hf} we show a 
schematic picture of the interactions involved in this model.
\begin{table}[t]
\centering
\caption{Parameters used for the mode hydrogen-bonded linear chain. All
quantities expressed in a.u.}\label{param}
\begin{tabular}{ccccccccccccccc}
\hline\hline
$D$ & &$b$ & &$r_e$ && $k_{FF}$& & $k_{HH}$&& $D_l$&& $b_l$&&$a_l$\\
\hline
0.0171 & &4.1276 && 1.7763 && 0.02351 && 0.06413&& 0.80848 && 0.3&& 4.4\\
\hline\hline
\end{tabular}
\end{table}

\begin{figure}[ht]
  \centering
  \includegraphics[scale=0.3,angle=270]{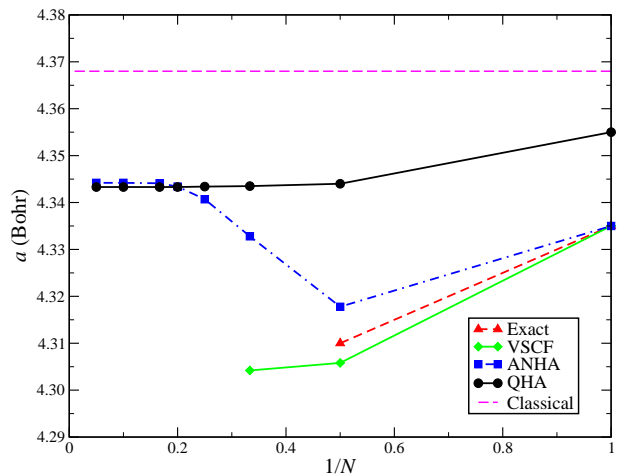}
  \caption{(Color online) Quantum-corrected lattice constant for a model H-bonded chain 
           as a function of the inverse number of cells. Contrary to covalently-bonded 
           systems, the quantum-mechanical effect is to decrease $a$ from its classical 
           value (horizontal long-dashed line on top). The green (light grey, solid) line 
           with diamonds represents the VSCF results, while the red (short-dashed) line with
           triangles are exact results. The blue (dot-dashed) line with squares is the
           ANHA, and the black (solid) line with circles is the QHA.}
  \label{a_vs_nc}
\end{figure}

Figure \ref{a_vs_nc} shows the equilibrium lattice parameter as a function 
of the inverse number of cells. The behavior of the QHA with the number of 
cells is similar to the monoatomic chain (black circles, solid line). Notice that 
here each cell contains 2 atoms, so that a well-converged value of $a_{QHA}$ requires at 
least four cells, i.e. 8 atoms. However, three cells are already quite well converged, 
and two cells produce a very reasonable value. Therefore, since this is an easily 
affordable size, in this paper we present results mostly obtained with two cells. 

Notice that now $a_{QHA}$ decreases with increasing $N$. In fact, 
at variance with the covalently bonded chain, the QHA converged value 
is 0.025 Bohr smaller than the classical value.
This is a characteristic 
feature of H-bonded systems, where the stretching frequency that contributes to the 
ZPE initially decreases upon compression because the H-bonds weaken. This occurs until 
the bond becomes symmetric, and only then the frequencies (and the ZPE) start to increase,
as will be shown in Section \ref{first_principles}.

Next, the normal modes determined at the energy minimum were used to compute the ZPE in
the anharmonic approximation (ANHA, blue squares, dot-dashed line). The converged 
value of the 
lattice constant is
essentially indistinguishable from $a_{QHA}$. Its behavior with $N$, however, is 
non-monotonic. 
This can be explained through the fact that, for this linear 
chain, there is only one vibrational coordinate that exhibits a double-well potential. 
As the number of cells is increased, the contribution of this double-well to the total 
energy becomes increasingly unimportant. 

In order to evaluate the effect of mode-coupling anharmonicities, we carried out VSCF 
calculations by mapping the PES along the normal modes corresponding to the minimum 
energy configuration, i.e. with all the H-atoms off-center (green diamonds, solid
line at the bottom). The exact calculation used the PES mapped along the normal modes 
corresponding to the saddle-point configuration, i.e. with all the H-atoms in the 
center of the bonds (red triangles, dashed line). 
At variance with the VSCF, the choice of vibrational coordinates is irrelevant because, 
unlike the case of molecular systems, the vibrational subspace in crystals does not
depend on the reference geometry. \cite{nikitas} 
In the VSCF approximation we went up to a 6-atom supercell 
($N$=3), i.e. a 5-dimensional VSCF problem. 
The good agreement of the VSCF with respect
to exact results for $N=2$, together with the trend with $N$ exhibited by the VSCF, suggest
that a large fraction of the quantum effect is already captured in the two-cell calculation. 
Nevertheless, this statement must be taken with caution. Calculations on
larger supercells are needed to verify this, especially in view of the unusual
behavior of the anharmonic approximation ANHA.

Another conclusion is that the QHA underestimates the lattice contraction due to quantum
effects, remaining at about half the required value. The next level of difficulty
and computational cost is the ANHA, but its non-monotonic behavior with system size 
makes it rather unsafe as a method to include anharmonicity. In particular, the
seemingly good results obtained for two cells appear to be fortuitous. The VSCF,
however, appears to be quite a good and robust method that reproduces exact results
to an excellent accuracy. This indicates that the correlation between modes is a
relatively minor effect, so that a mean-field approximation where each mode feels
the other modes on average, is sufficient for this class of problems.


\subsection{First-principles hydrogen-bonded chain\label{first_principles}}

As a realistic application we studied an F-H hydrogen-bonded chain by means of first-principles 
calculations. Actual F-H chains develop a zig-zag structure. \cite{sankey} However, since our 
purpose here is not to solve the problem of F-H crystals, but to understand isotope effects when 
a realistic PES is used, we disregard this geometrical aspect and consider straight chains and 
the motion of the atoms only along the axis of the chain. First-principles calculations have 
been carried out using a combination of codes. For phonon calculations we used the 
pseudopotential plane wave code Quantum-ESPRESSO. \cite{espresso} The energy curves and 
lattice constants were calculated with SIESTA, \cite{siesta} which also uses pseudopotentials 
but the wave functions are expanded in a localized basis set of pseudo-atomic orbitals. 
All calculations were carried out within the PBE generalized gradient approximation 
\cite{pbe} to density functional 
theory (DFT). 

\begin{figure}[ht]
  \includegraphics[scale=.3,angle=270]{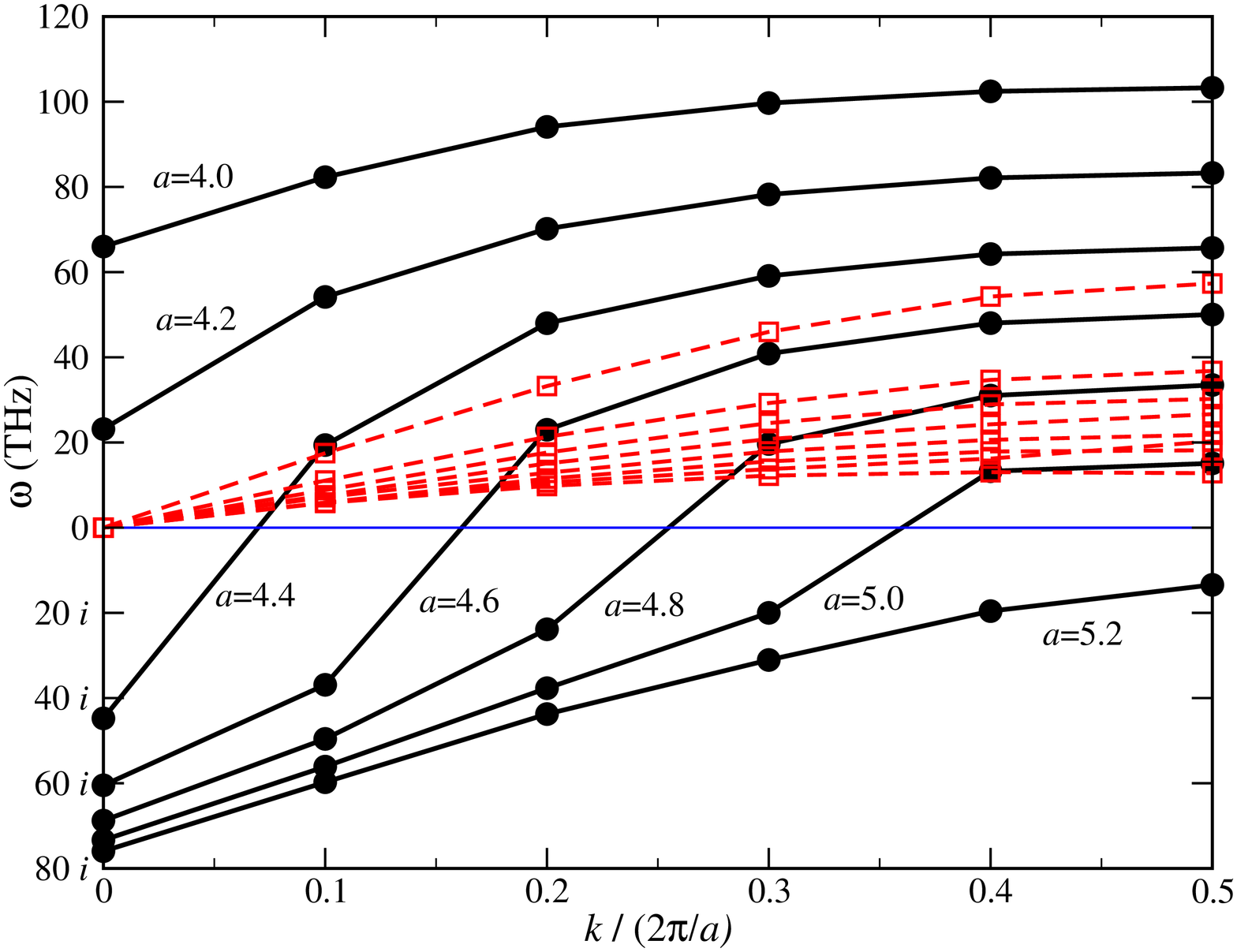}
  \includegraphics[scale=.3,angle=270]{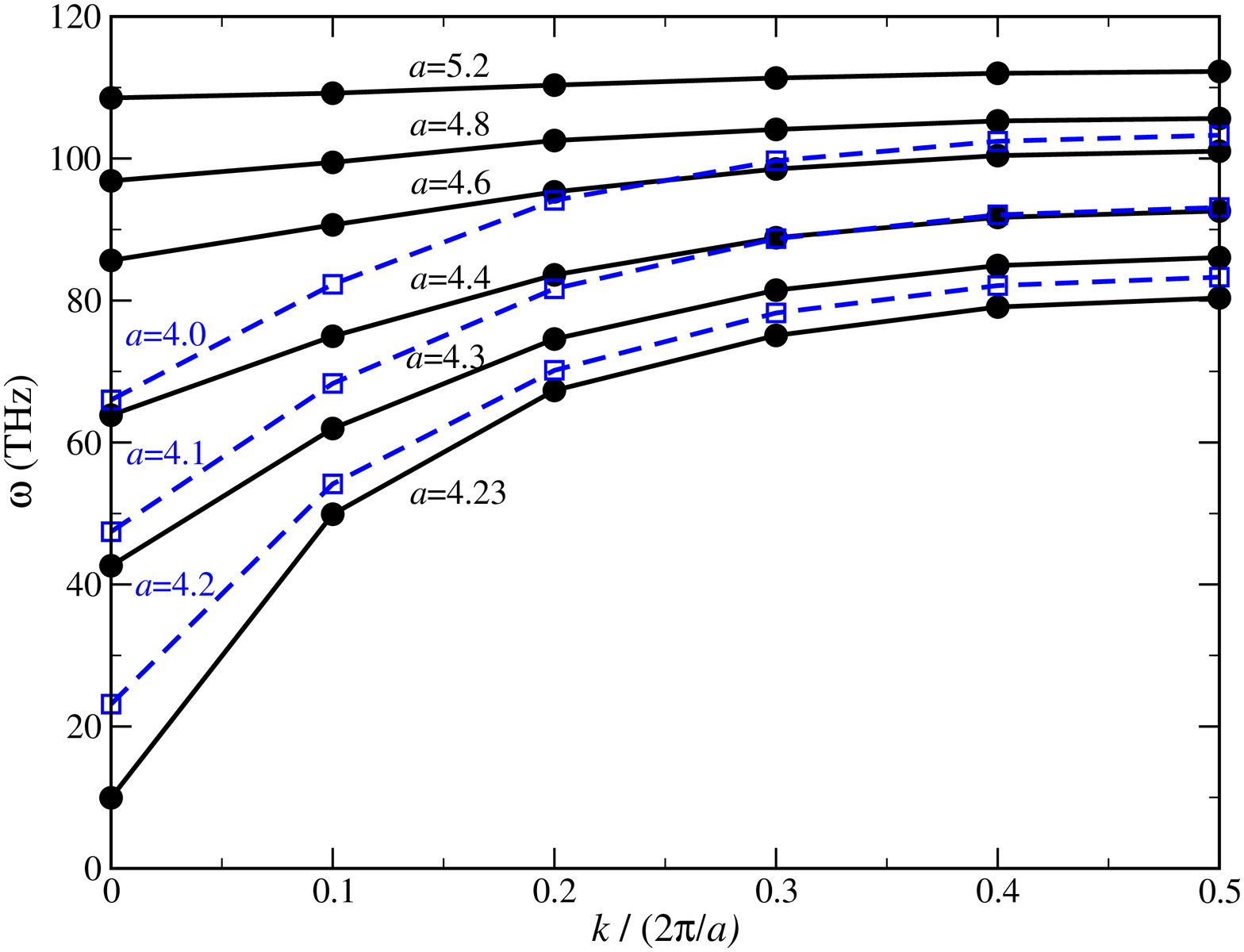}
  \caption{(Color online) First-principles phonon dispersions for a linear F-H chain at 
           various lattice parameters. Upper panel: H-atoms in centered positions. Red 
           (dashed) lines with open squares represent acoustic branches, while black 
           solid lines with filled circles correspond to optical branches. Lower panel: 
           H-atoms in their stable position. Blue (dashed) lines with open squares 
           correspond to stable centered H-atoms, while black solid lines with filled 
           circles indicate stable off-centered H-atoms. Acoustic branches are omitted 
           for clarity. Values of lattice constants in Bohr are indicated in the figure.}
  \label{fh_phonon}
\end{figure}
In Fig. \ref{fh_phonon} (upper panel) we present the phonon dispersion relations along 
the direction of the chain, calculated in the symmetric configuration with the H-atoms 
in their centered position, from a compressed lattice constant of 4 Bohr up to an 
expanded value of 5.2 Bohr. For small $a$ there is no double-well, the H-atoms 
are stable in their centered positions and the optical phonon branch for this reference 
configuration is stable. At $a\approx 4.22$ Bohr the double-well emerges and the 
H-atoms prefer an off-centered configuration. If we insist on calculating the phonon
dispersions for the centered configuration, the optical branch develops an instability
at zone-center. In fact, there is a whole portion of the optical branch that is unstable.
The extent of the instability region increases with increasing lattice constant, to the 
point of making the optical mode unstable throughout the whole Brillouin zone when the 
chain is stretched to 5.2 Bohr. 

Further insight is obtained by calculating the phonon dispersion relations for the stable
configuration. These are presented in the lower panel of Fig. \ref{fh_phonon}, where 
it can be observed that the zone-center optical phonon becomes soft at $a\approx 4.22$ Bohr, 
and after the H-atoms have centered (blue lines and open squares) its frequency starts 
rising again quite steeply. An interesting observation is that the curvature of the optical 
phonon branch is the opposite of the usual picture where the frequency decreases from
zone center towards zone boundary. This is a consequence of the type of bonding, and common
to hydrogen-bonded systems. In the zone-center optical phonon all unit cells move in phase,
with one H-atom approaching the F-atom while the other moves away. At zone boundary, 
consecutive unit cells move out of phase so that half of the F-atoms are approached by two 
H-atoms rather than one. This is in clear violation of the ice-rules, which penalize this
type of motion with an increased energy, thus explaining why the optical mode is harder 
at zone-boundary.

\begin{figure}[ht]
  \centering
  \includegraphics[scale=0.30,angle=270]{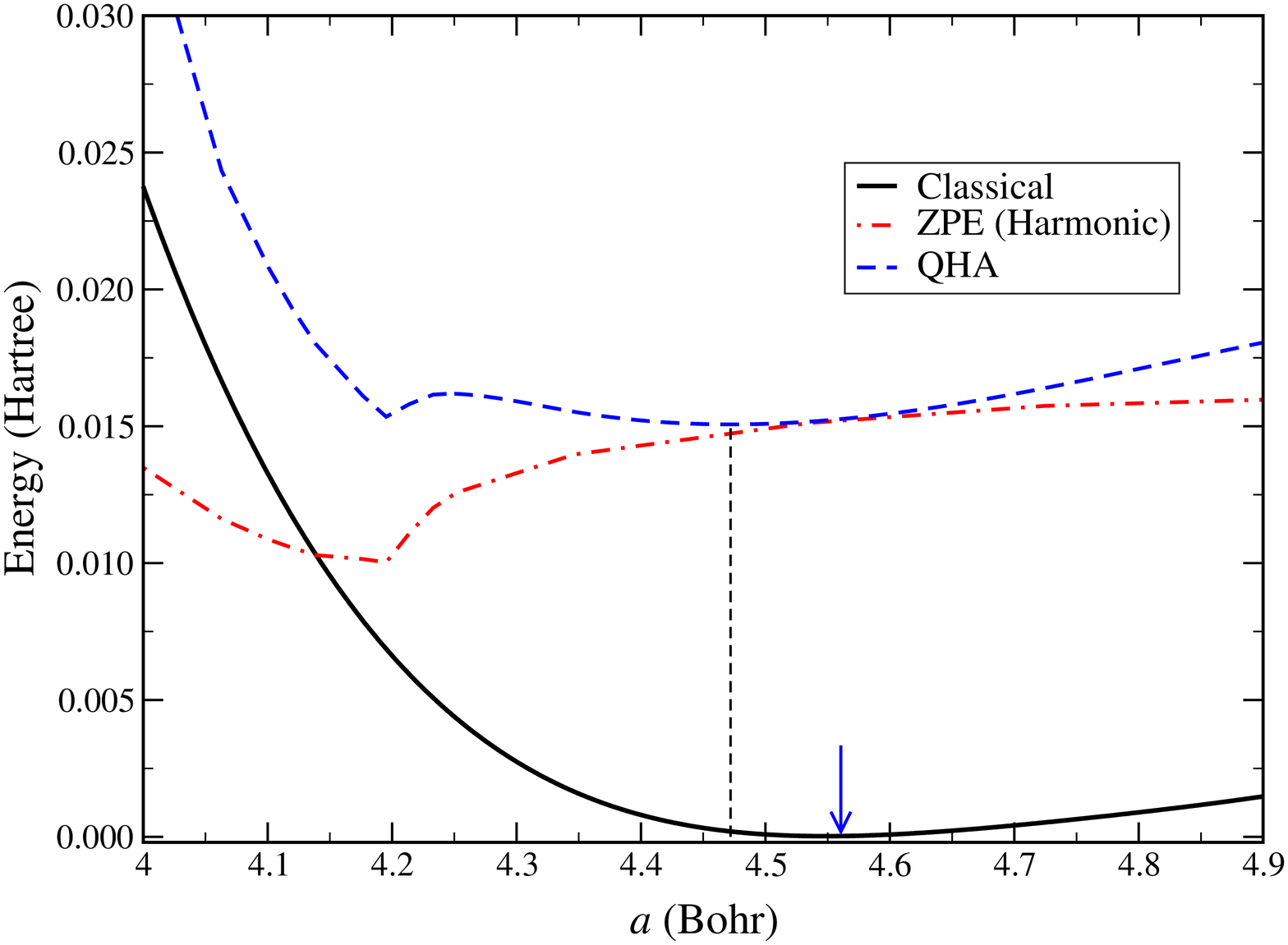}
  \includegraphics[scale=0.30,angle=270]{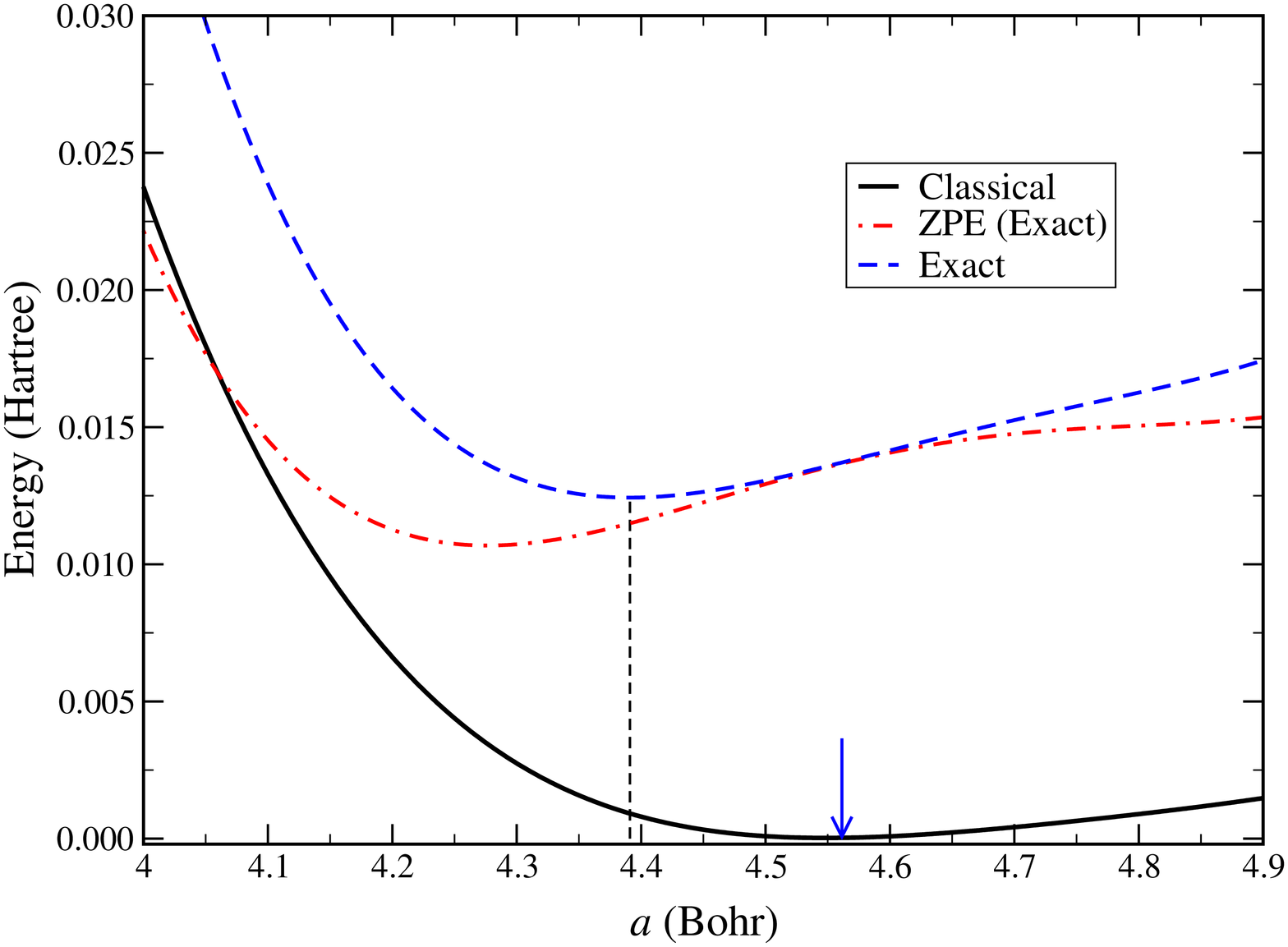}
  \caption{(Color online) Upper panel: Harmonic ZPE (red, dot-dashed line) and QHA 
           corrected energy curve (blue, dashed line). Lower panel: exact ZPE and 
           quantum-corrected energy as a function of lattice parameter. Color and
           line coding as in upper panel. The classical energy curve (black, solid 
           line) is shown in both panels for comparison.
           Note the large difference between QHA and exact equilibrium lattice
           constant (vertical dashed lines). Arrows indicate the classical lattice
           constant.}
  \label{zpe_abinitio}
\end{figure}

The harmonic frequencies can now be used to analyze the behaviour of the QHA as a function 
of the lattice constant. 
In Fig. \ref{zpe_abinitio} (upper panel) we show the ZPE 
(red, dot-dashed line), the classical PES (black, solid), and the quantum-corrected PES 
(blue, dashed) as a function of the lattice parameter for $N=2$ (4 atoms). It is interesting 
to analyze the behavior of the QHA when the system is highly compressed.
It is apparent that the ZPE exhibits a non-analytic 
behavior (a cusp) at $a\approx 4.2$ Bohr, where the H-atoms center and the type of bonding 
changes from hydrogen-bonding to covalent. As a consequence, the QHA curve exhibits two minima, 
one of which is spurious. Since this characteristic is associated to the stretching mode, and
the remaining modes are not severely affected by compression, similar behaviour is observed
for a larger number of cells. This feature represents a physically incorrect picture, and 
thus a clear limitation of the QHA in the description of hydrogen-bonded systems at high 
pressures. \cite{sankey} In fact, this non-analytic behaviour completely disappears when 
the problem is solved exactly (lower panel). Notice the substantial difference (0.07 Bohr) 
between the QHA and exact lattice constants (dashed vertical lines).

Similarly to the model presented in Section \ref{diatomic}, we have examined the dependence
of quantum nuclear effects on the number of cells, and for the various approximations. 
Results are summarized in table \ref{a_vs_app_nc_hf} below:
\begin{table}[ht]
\vspace{-0.05cm}
\centering
\caption{Equilibrium lattice parameters (in Bohr) at various levels of approximation.}
\label{a_vs_app_nc_hf}
\begin{tabular}{ccccccc}
\hline\hline
 No. cells && Classical & QHA & ANHA & VSCF & Exact \\
\hline
1 && 4.560 & 4.195 & 4.381 & 4.381 & 4.381 \\
2 && 4.560 & 4.462 & 4.387 & 4.390 & 4.392 \\
\hline\hline
\end{tabular}
\end{table}

\begin{figure}[ht]
\centering
  \includegraphics[scale=0.3,angle=270]{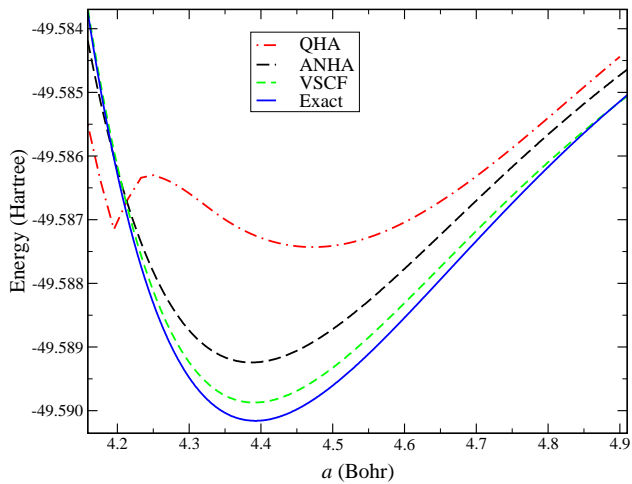}
  \caption{(Color online) Ground state energy as a function of lattice parameter for two 
            unit cells at various levels of approximation: The blue (solid) line is for
            the exact calculation, the green (dashed, light grey) is the VSCF, the
             black (long-dashed) is the ANHA, and the red (dot-dashed) the QHA.}
\label{2_cell_ab}
\end{figure}

For a single cell there is only one 
vibrational coordinate associated to the F-H stretching, which is subject to a double-well 
potential. Since there is only one mode, both ANHA and VSCF approximations are equivalent 
to the exact solution. The lattice constant obtained in the QHA, however, is anomalously
small.
By inspection it is clear that this value corresponds to the spurious minimum mentioned 
above, which for one cell appears to be lower in energy than the correct minimum.
Figure \ref{2_cell_ab} shows the quantum-corrected energy as a function of lattice 
parameter for two cells (i.e. two F-H units, three vibrational modes). 

At variance with a single cell, the QHA (red, dot-dashed line) provides a 
better estimate of the effect.  Using the vibrational coordinates obtained in the stable 
configuration, and already used for the QHA, we introduced intra-mode anharmonicities 
through the ANHA (black, long-dashed), and mode-coupling anharmonicities via the 
VSCF approach (green, dashed -- light grey in print). The latter is in very good agreement 
with the exact result (blue, solid), especially for small and large values of the lattice 
parameter. 

\begin{figure}[ht]
\centering
  \includegraphics[scale=0.3,angle=270]{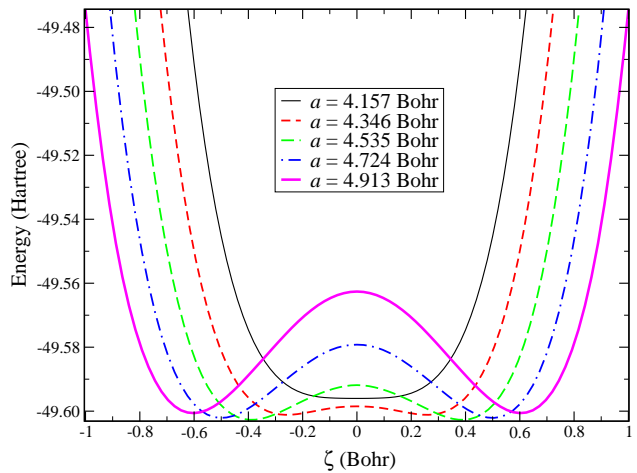}
  \caption{(Color online) First-principles PES along the zone-center optical mode for 
           different values of the lattice parameter. The thin black curve represents
           the most compressed situation at $a=4.157$ Bohr. The red curve (dashed) 
           corresponds to $a=4.346$ Bohr, the green (long-dashed, light grey) to 
           $a=4.535$ Bohr (close to the equilibrium value), the blue (dot-dashed) to
           $a=4.724$ Bohr, and the magenta (thick solid, dark grey) curve to an expanded 
           lattice constant of $a=4.913$ Bohr. Energies are reported for a supercell 
           containing two unit cells. }
  \label{pot_fe_hf}
\end{figure}

In order to assess the quality of the various approximations, we mapped the PES along 
the zone-center optical mode for different values of the lattice parameter (Fig. 
\ref{pot_fe_hf}). For small values of $a$ the H-atoms are centered and subject to a 
single-well, anharmonic potential (black, solid thin line in Fig. \ref{pot_fe_hf}). 
Therefore, at high compression the QHA is a good approximation, but closer to the 
de-centering point the potential becomes more 
anharmonic, thus compromising its quality. This situation is notably improved by introducing 
intra-mode anharmonicities in the approximation of non-interacting modes (ANHA). The VSCF 
approximation reproduces the exact results to an excellent extent, thus indicating that the 
effect of correlation between modes is very small. A similar trend is observed for large 
values of the lattice parameter. In this case, the barrier in the double-well potential is 
so high that the overlap between the two degenerate states (left and right) is very small. 
Again, the ANHA provides an improved estimate of the energy with respect to the QHA, and 
the VSCF reproduces quite well the exact results, thus indicating that correlation between 
modes remains negligible. We conclude that in the two limiting cases the ANHA is a rather 
decent approach, and the VSCF reproduces very closely the exact results.

The situation is somewhat different at intermediate values of the lattice parameter, where 
the barrier is low and the effect of anharmonicity is larger. Here the QHA and ANHA are not 
very satisfactory, and have to be improved by introducing mode-coupling anharmonicities at the mean-field level, as in the VSCF scheme.
It is also in this region where correlations become more important. 

Although the energy curves are shifted,
the position of the minimum estimated by the ANHA is very good. According to the results
presented in Section \ref{diatomic} for the model hydrogen-bonded chain, we believe this 
may be a fortuitous coincidence for this particular system size. 
The QHA, however, is evidently not sufficiently accurate.

\subsection{Choice of vibrational coordinates in the VSCF}

The quality of the VSCF approximation depends on the choice of vibrational coordinates. 
The more uncorrelated they are, the better the approximation. Normal modes are a good starting 
point because in the harmonic limit they are completely uncorrelated. However, as the amplitude 
of the displacements along some {\it soft} modes increases (due to quantum delocalization), 
the normal modes begin to couple. We now analyze the question of which
vibrational coordinates are optimal for the VSCF approximation, in the case of the 4-atom cell.
The three modes are the zone-center optical phonon, which exhibits a double well, and the 
zone-boundary acoustic and optical phonons, which are practically harmonic. In principle, 
any linear combination of these three modes is possible. Nevertheless, since the modes at 
different {\bf k}-vectors do not mix, we fix the double-well mode and optimize the choice 
of modes in the two-dimensional zone-boundary subspace. \cite{notice}

\begin{figure}[ht]
\centering
  \includegraphics[scale=0.3,angle=270]{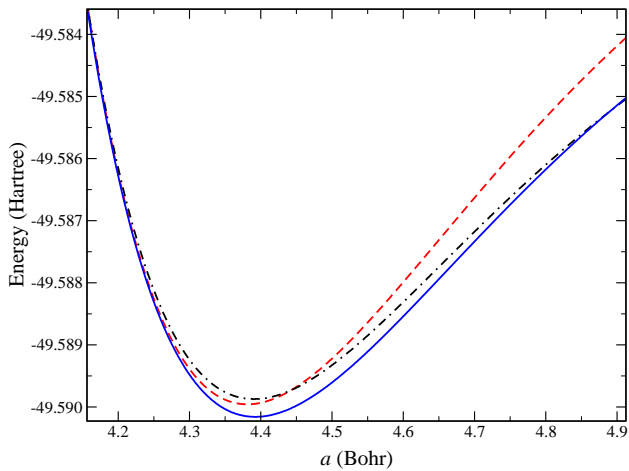}
  \caption{(Color online) Comparison of the VSCF calculations for two different choices 
           of vibrational coordinates. The red (dashed) line is for the saddle-point 
           modes, while the black (dot-dashed) line is for the modes calculated at the 
           classical equilibrium configuration. The blue, solid line represents the exact 
           energy.}
  \label{vscf_ex_hf}
\end{figure}

We studied the quality of the VSCF 
for two different choices: the normal modes calculated at the minimum of the PES, and those 
calculated at the saddle point configuration, with the H-atoms centered. The VSCF energies 
are reported in Fig. \ref{vscf_ex_hf} together with the exact energy. According to Fig. 
\ref{pot_fe_hf}, the two minima in the double-well get closer and eventually merge, as the 
lattice parameter is reduced. Therefore, for values of $a$ where the double-well has 
disappeared, there is no saddle point, and the normal modes at the minimum are an excellent 
choice for the VSCF. Similarly, at large $a$ the barrier separating the two minima is high, 
and again the wave function is quite localized around the two minima. 

Therefore, the vibrational coordinates calculated at the minimum of the PES are generally 
better than the saddle-point modes for solving the VSCF problem. At intermediate values 
of $a$ none of the two choices is clearly superior to the other. We have tried to optimize 
the choice of modes within this subspace by minimizing the ground state VSCF energy, but the energy gain 
was always of the same order of magnitude of the difference between minimum and saddle-point 
modes. Therefore, this appears to be the limit of the VSCF. Any further improvement requires 
the introduction of correlation between modes. The present situation is reminiscent of static 
correlation cases often encountered in electronic structure calculations, where a single 
Slater determinant (an uncorrelated electronic configuration) is insufficient to represent
the ground state. These require a multi-determinantal wave function as in multi-reference
methods like CASSCF. Here, a linear combination of two Hartree products, each one 
corresponding to one of the two equivalent minima, constitutes an improved wave function 
that correctly respects the symmetry of the problem. It is well-known that static correlation 
cannot be recovered by perturbative methods based on a single reference.\cite{norris} This is
also a problem for truncated CI expansions, while full-CI would require a large basis set. 
This is why, in the field of molecular spectroscopy, various groups have developed 
multi-configuration methods such as MCTDH.\cite{mctdh}

\subsection{Isotope effects\label{isotope}}

\begin{figure}[ht]
  \includegraphics[scale=0.3,angle=270]{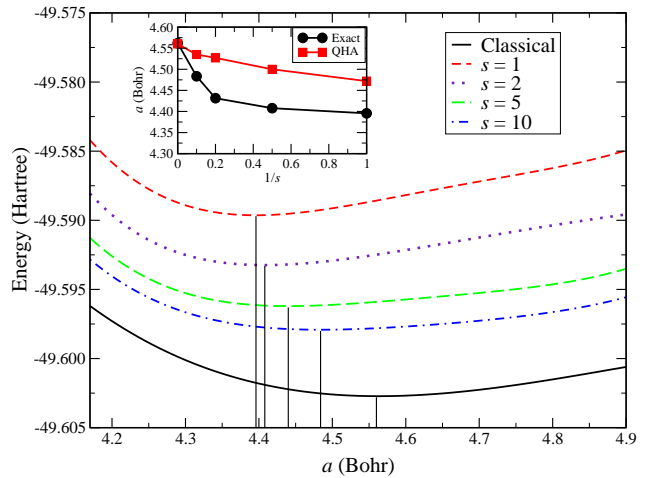}
  \caption{(Color online) Energy vs. lattice constant for various mass-scaling factors.
           The black (solid) line is the classical result, blue (dot-dashed) for $s=10$,
           green (long-dashed, light grey) for $s=5$, violet (dotted) for $s=2$, and red
           (short-dashed, dark grey) for $s=1$. Vertical lines 
           indicate the equilibrium lattice constants, which are then reported in the 
           inset as a function of $1/s$. The HF chain PES has been calculated at the 
           DFT-PBE level, using the SIESTA code. Actual HF chains assume a zig-zag form, 
           but here we considered linear chains.}
  \label{isot_hf}
\end{figure}

\begin{figure}[ht]
  \includegraphics[height=.22\textheight]{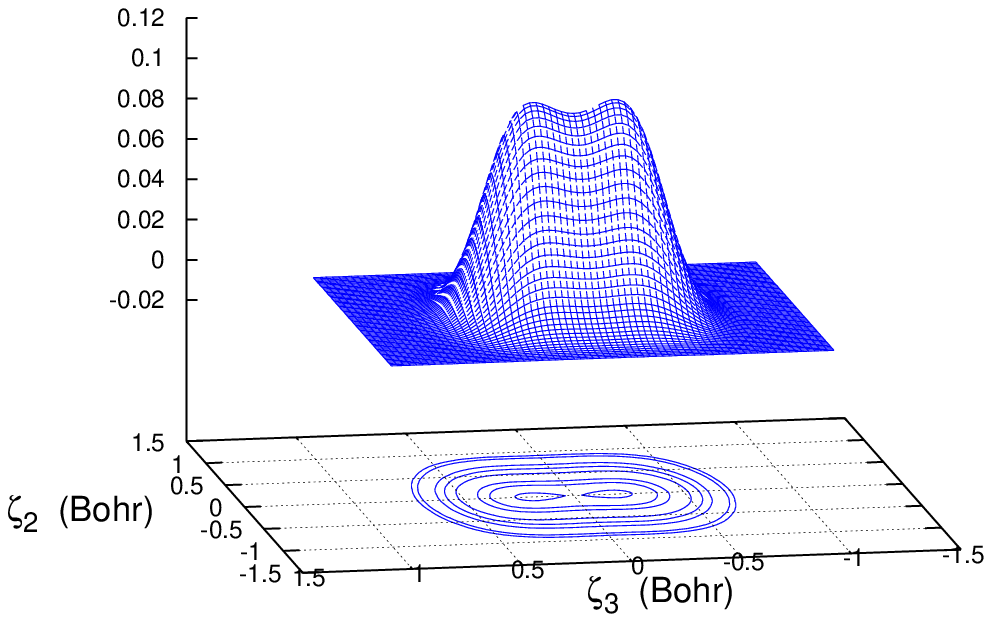}
  \includegraphics[height=.22\textheight]{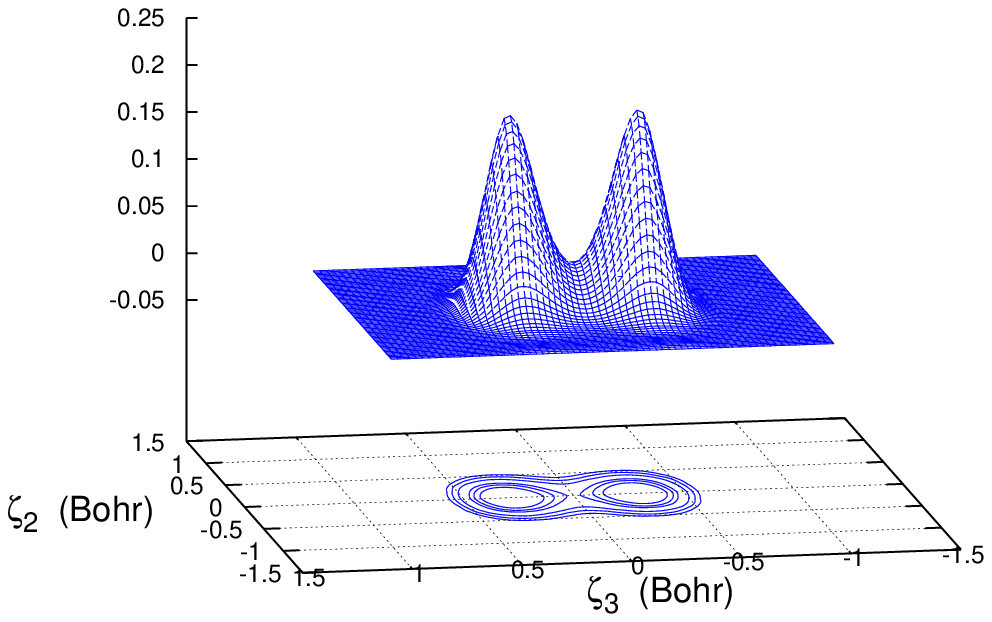}
  \caption{Two-dimensional cut of the ground state wave function in the subspace of the 
           two optical modes $\zeta_2$ and $\zeta_3$. The upper panel is for the true 
           masses and the lower panel for particles five times heavier. Notice the enhanced 
           localization of the latter.}\label{wf_hf}
\end{figure}

Figure \ref{isot_hf} shows the effect of modifying the nuclear masses on the lattice 
parameter for the case of two F-H cells. In order to evaluate isotope effects, both 
the H and F masses were multiplied by a scale factor $s$, so that the mode eigenvectors,
and thus the vibrational coordinates, were not modified.
We have then solved the three-dimensional Schr\"odinger equation in the vibrational space 
for various lattice constants, and obtained the quantum corrected energy curves for the 
true H and F masses (red, short-dashed), when these masses are multiplied by two (violet, 
dotted), five (green, long-dashed), and ten (blue, dot-dashed). For comparison we also show 
the classical energy curve, corresponding to the limit of infinite masses. As expected, 
the energy increases with decreasing mass, while the lattice constant decreases, as 
illustrated in the inset. This is the essence of the geometric or Ubbelohde effect in H-bonded 
systems. \cite{ubbelohde} Lighter particles are more delocalized and exhibit a larger 
probability in the region of the barrier. According to Fig. \ref{pot_fe_hf}, this favors 
more compressed bonds, thus translating into an effective attraction between the two 
neighboring F-atoms (or oxygens in the more common O-H$\cdots$O hydrogen bonds). This 
attraction is reflected in an enhanced cohesion, and thus a smaller lattice constant, 
\cite{kdp} which is a common feature of a large family of H-bonded crystals. \cite{ferro}
To illustrate this concept of enhanced delocalization, we show in Fig. \ref{wf_hf} a 
two-dimensional cut of the wave function for the true masses and those multiplied by 
five, for a lattice constant $a$=4.44 Bohr.

We can now use these wave functions to calculate the quantum expectation value of the F-H 
distance as a function of the mass of the particles. This is a very important quantity, 
because it is what is obtained in neutron and X-ray diffraction experiments. This distance is 
usually quite different from that obtained classically for the same lattice parameter, which 
corresponds to the minimum of the double-well PES. In fact, this is probably the source of many 
inconsistencies between experimental structures and those determined via first-principles 
calculations. \cite{kdp} In Table \ref{d_fh_vs_s} we report the exact F-H distance as a 
function of the mass scaling factor
for a lattice constant of 4.44 Bohr.
\begin{table}[!h]
\centering
\caption{Equilibrium F-H distance (in Bohr) as a function of the mass scaling factor $s$
         for a lattice constnat of 4.44 Bohr.}
\label{d_fh_vs_s}
\begin{tabular}{ccccccc}
\hline\hline
$s$ &&  1  &  2  &  5  &  10  &  $\infty$ \\
\hline
$d_{\rm OH}$ && 2.08 & 2.06 & 2.02 & 1.99 & 1.85 \\
\hline\hline
\end{tabular}
\end{table}
As expected, the F-H distance decreases when the masses become heavier. 
The classical F-H distance increases by a substantial 10\% 
when the true H and F masses are used.
In contrast, at fixed lattice constant the QHA 
values are insensitive to $s$. This is because quantum and classical harmonic oscillators have 
the same equilibrium position, and thus the quantum harmonic approximation does not modify 
the internal geometry. More generally, the F-H distance in the QHA corresponds to the 
classical value obtained at the $s$-dependent equilibrium lattice constant, which is
reported in the inset to Fig. \ref{isot_hf} (red symbols). Therefore, the influence of 
mass-scaling is limited to the {\it volume} effect, which is significantly smaller than the 
full quantum effect. Firstly, the dependence of the QHA lattice constant with $s$ is milder
than the exact one.
Secondly, as shown
in Fig. \ref{pot_fe_hf}, variations of this magnitude lead to small changes in the 
location of the minima of the double-well. As a consequence, the F-H distance is severely
underestimated in the QHA, thus precluding its use in the description of the internal 
geometry of H-bonded systems.

In the present case, the isotope effect upon doubling of all the masses, which is closely related 
to deuteration, does not entail a significant change in the F-H distance. In a first instance 
one could think that this is because it was calculated at fixed lattice constant (see table 
\ref{d_fh_vs_s}). In effect, this would be consistent with theoretical calculations \cite{kdp} 
and experiments under pressure. \cite{nelmes} When sufficient pressure is applied to deuterated 
compounds to reproduce the lattice parameters of its protonated analogue, the internal 
geometries turn out to be quite similar. Within this context, one of us has shown that most 
of the isotope effect arises from a self-consistent interplay between wave function 
localization, internal geometry and lattice parameters. \cite{kdp_prl} 
Therefore, to assess the full extent of the isotope effect one would have to compare 
the internal geometries at the corresponding equilibrium lattice constants. 
In the present case we
obtained F-H distances of 2.12 and 2.10 Bohr, respectively. Interestingly, the isotope
effect remains as small as before. This is not inconsistent with experimental data,
though. In fact, the magnitude of the isotope effect on inter-atomic distances in 
hydrogen-bonded systems ranges from almost insignificant values for some systems, to large 
differences as in KDP, where the O-H distance of 2.45 \AA~rises to 2.52 \AA~in 
its deuterated analogue DKDP. \cite{nelmes} This depends on several factors, mainly
the size of the coupling between internal coordinates and strain and the shape of
the double-well (barrier height and distance between minima) in the region around the 
equilibrium lattice constant.

\section{Conclusions and outlook\label{conclusions}}

We have analyzed the influence of quantum nuclear effects on the structural 
properties of solids. To this end we have
implemented a simple methodology of mapping the multidimensional PES in the 
space of normal modes, and then solving the resulting vibrational 
Schr\"odinger equation. This has been done exactly when possible, and  also
using a number of computationally tractable approximations that range from the 
vibrational self-consistent field (VSCF) method and its variants, down 
to the quasi-harmonic approximation (QHA).

We have studied the behaviour of energies and lattice constants in
an anharmonic monoatomic chain as a function of the size of the supercell, 
and showed that results for four cells are probably already 
quite close to convergence. Here, the QHA is an excellent approximation unless
the masses become exceedingly small, thus justifying the customary approach
followed in solid-state physics. We have then analyzed the behaviour of a 
model hydrogen-bonded chain, and observed that the coupling between modes can be
important. Finally, we considered a realistic H-bonded chain where the
PES was determined from first-principles calculations. Here we obtained
the quantum-corrected lattice constant and internal structural parameters
(distances), and showed how isotope effects arise in these systems. We have
also analyzed the various approximations and showed that the QHA is 
insufficient, while the independent-mode anharmonic approximation ANHA 
appears to introduce an important improvement at the energetic and structural
level. Nevertheless, results obtained for the model hydrogen-bonded chain
suggest that this may be accidental and due to size effects.
VSCF results are very close to exact ones, thus suggesting that correlation
between modes is a minor effect. Optimizing the vibrational coordinates in 
the VSCF method does not constitute a significant advantage; further 
improvement should arise from correlated methods such as vibrational CI or
MCTDH. 
In any case, the VSCF approach using the modes calculated at the classical 
equilibrium geometry appears to be a very good approximation.

Although technically the extension to higher dimensionalities is straightforward, 
there are some features that are not present in one-dimensional systems. 
For example, the tunneling mode can be strongly coupled to other modes
\cite{giuseppe_thesis} so that, for large amplitude of motion, correlation 
between them may become important. In this case we enter the realm of 
multi-dimensional tunneling where, if we insist on describing the proton motion
in terms of a single, effective tunneling coordinate, this latter tends to be
curvilinear. Therefore, it is quite likely to have to adopt schemes
that combine the exact solution for one or a few subspaces of strongly coupled 
modes, with a VSCF representation or a lower-level scheme such as the QHA for 
the remaining modes. \cite{vendrell} It remains the problem of identifying which 
are those subspaces, though.

There are many aspects of this approach that can be improved, especially 
for what concerns simplified methods that can be used for larger systems
and higher dimensionality. Within the VSCF scheme, the limiting factor is
the mapping of the PES and the multi-dimensional integration. The latter
can be efficiently dealt with by factorizing the PES into a sum of products 
of modes. \cite{product} This allows for the computation of VSCF potentials 
as products of inexpensive one-dimensional integrals. The problem of fitting the PES 
remains open, although there are general strategies for PES fitting based on 
interpolation methods, \cite{rabitz1} or using principles of multivariate 
analysis such as the high-dimensional model representation of Rabitz and 
Ali\c{s}. \cite{rabitz2} An improvement to the ANHA that approximately recovers 
the interaction between modes at the mean-field level consists of
using the quantum expectation value of the geometry as a reference, rather
than the classical one. This can also be seen as a simplification of the
VSCF approximation where the vibrational wave functions in the integrals are
replaced by completely localized delta-functions. This is a good approximation
unless one or more approximated wave functions exhibit two peaks corresponding 
to a double-well potential. 

Another subtle issue in hydrogen-bonded systems is that, although DFT-GGA
approaches reproduce structural properties quite well, proton transfer barriers 
appear to be exceedingly small, in some cases even disappearing altogether.
\cite{scuseria} Therefore, other electronic structure methods such as hybrid 
Hartree-Fock-DFT \cite{truhlar2} or correlated quantum chemical methods have to be 
explored. \cite{manby}

A straightforward extension of this methodology is the calculation of
vibrational excitations. Once the PES is known, excited states can be
easily calculated and used to compute thermodynamic quantities beyond 
the quasi-harmonic approximation.
General VSCF algorithms useful to tackle mild anharmonicities in molecular systems 
have been implemented many years ago, \cite{adrian} and more recently in conjunction
with an ab initio description of the PES. \cite{chaban} In fact, a VSCF option is 
available in some electronic structure codes like GAMESS. \cite{gamess} The main
use of this capability is at the spectroscopic and thermochemical, rather than 
structural level.
An extension of this methodology to crystalline systems is straightforward and 
perfectly viable, although to the best of our knowledge this has not been implemented
so far. Nevertheless, the treatment of highly anharmonic systems such as those 
involving double-wells remains a challenge. We expect the present paper to be a
relevant contribution in this direction.

We thank David Hughes and Alfredo Caro for important contributions in the
initial stages of this project. We also thank Pietro Ballone, Mario Del Popolo 
and Ricardo Migoni for helpful discussions. 

\vfill


\begin{thebibliography}{9}

\bibitem{qha1} G. Liebfried and W. Ludwig, {\it Solid State Physics} (Academic New York, 1961), vol. 12, page 276.
\bibitem{qha2} R. A. Cowley, Adv. Phys. {\bf 12}, 421 (1963).
\bibitem{qha3} L. N. Kantorovich, Phys. Rev. B {\bf 51}, 3520 (1995).
\bibitem{kittel} C. Kittel, {\it Introduction to solid state physics} (Wiley, NY, 1986).
\bibitem{scivetti_aip} I. Scivetti, D. Hughes, N. I. Gidopoulos, A. Caro, and J.
Kohanoff, AIP Conference Proceedings {\bf 963}, 212 (2007).
\bibitem{giuseppe_thesis} G. Colizzi, PhD thesis (Queen's University Belfast, 2005)
\bibitem{baroni_01} S. Baroni, S. de Gironcoli, A. Dal Corso, and P. Giannozzi, 
Rev. Mod. Phys. {\bf 73}, 515 (2001).
\bibitem{product} A. J\"ackle and H.-D. Meyer, J. Chem. Phys {\bf 104}, 7974 (1996); 
{\it ibid} {\bf 109}, 3772 (1998).
\bibitem{watts} M. A. Suhm and R. O. Watts, Phys. Rep. {\bf 204}, 293 (1991).
\bibitem{baye} D. Baye and P.-H. Heenen, J. Phys. A {\bf 19}, 2041 (1986); D. Baye, Phys.
Stat. Sol. (b) {\bf 243}, 1095 (2006).
\bibitem{varga} K. Varga, Z. Zhang, and S. T. Pantelides, Phys. Rev. Lett. {\bf 93}, 176403 (2004).
\bibitem{gerber} M. A. Ratner and R. B. Gerber, J. Phys. Chem. {\bf 90}, 20 (1986).
\bibitem{adrian} A. Roitberg, R. B. Gerber, R. Elber, and M. A. Ratner, Science
{\bf 268}, 1319, (1995).
\bibitem{truhlar} T. C. Thompson and D. G. Truhlar, J. Chem. Phys. {\bf 77}, 3031 (1982).
\bibitem{gallegos} A. Hidalgo, J. Z\'u\~niga, J. M. Franc\'es, A. Bastida and A. Requena, 
Int. J. Q. Chem., Vol. 40, 685-694 (1991).
\bibitem{yanovitskii}  O. Yanovitskii, G. Vlastou-Tsinganos, and N. Flytzanis, Phys. Rev. B 
{\bf 48}, 12645 (1993).
\bibitem{nikitas} I. Scivetti, PhD Thesis (Queen's University Belfast, 2008).
\bibitem{sankey} R. W. Jansen, R. Bertoncini, D. A. Pinnick, A. I. Katz, R. C. Hanson, 
O. F. Sankey, and M. O'Keeffe, Phys. Rev. B {\bf 35}, 9830 (1987).
\bibitem{espresso} P. Giannozzi et al., http://www.quantum-espresso.org.
\bibitem{siesta} J. M. Soler, E. Artacho, J. Gale, A.  Garc\'{\i}a, J. Junquera, P. Ordej\'on,
and D. S\'anchez-Portal, J. Phys.: Condens. Matter, {\bf 14}, 2745 (2002).
\bibitem{pbe} J. P. Perdew, K. Burke, and M. Ernzerhof, Phys. Rev. Lett. {\bf 77}, 3865 (1996).
\bibitem{notice} Notice that for larger supercells there will be coupling between zone-center 
optical modes. In the case of higher dimensionality, this coupling may introduce a curvature 
in the double-well coordinate that has to be treated carefully.
\bibitem{norris} L. Norris, M. A. Ratner, A. E. Roitberg, and R. B. Gerber, J. Chem. Phys. 
{\bf 105}, 11261 (1996).
\bibitem{mctdh} H.-D. Meyer, U. Manthe, and L. S. Cederbaum, Chem. Phys. Lett. {\bf 165}, 
73 (1990).
\bibitem{ubbelohde} J. M. Robertson and A. R. Ubbelohde, Proc. R. Soc. London A {\bf 170}, 222 (1939).
\bibitem{kdp} S. Koval, J. Kohanoff, J. Lasave, G. Colizzi, and R. L. Migoni, Phys. Rev. B {\bf 71},
184102 (2005).
\bibitem{ferro} R. Blinc and B. Zeks, in {\it Soft Modes in Ferroelectrics and Antiferroelectrics}, 
edited by E. P. Wohlfarth (North-Holland, Amsterdam, 1974).
\bibitem{nelmes} R. J. Nelmes, M. I. McMahon, R. O. Piltz, and N. G. Wright, Ferroelectrics 
{\bf 124}, 355 (1991). 
\bibitem{kdp_prl} S. Koval, J. Kohanoff, R. L. Migoni,  and E. Tosatti, Phys. Rev. Lett. {\bf 89},
187602 (2002).
\bibitem{vendrell} O. Vendrell, V. Gatti, V. Lauvergnat, and H.-D. Meyer, J. Chem. Phys. {\bf 127}, 
184302 (2007).
\bibitem{rabitz1} T.-S. Ho and H. Rabitz, J. Chem. Phys. {\bf 104}, 2584 (1996).
\bibitem{rabitz2} H. Rabitz and \"O. F. Ali\c{s}, J. Math. Chem. {\bf 25}, 197 (1999).
\bibitem{scuseria} S. Sadhukhan, D. Mu\~noz, C. Adamo, and G. E. Scuseria, Chem. Phys. Lett. 
{\bf 306}, 83 (1999).
\bibitem{truhlar2} B. J. Lynch and D. G. Truhlar, J. Phys. Chem. A {\bf 105}, 2936 (2001).
\bibitem{manby} L. Maschio, D. Usvyat, F. R. Manby, S. Casassa, C. Pisani and M. Schutz, Phys. Rev. B {\bf 76}, 075101 (2007).
\bibitem{chaban} G. M. Chaban, J. O. Jung, and R. B. Gerber, J. Chem. Phys. {\bf 111}, 
1823 (1999).
\bibitem{gamess} http://www.msg.chem.iastate.edu/GAMESS/GAMESS.html
\end{thebibliography}
\end{document}